\documentclass[aps, prb, twocolumn, superscriptaddress]{revtex4-2}
\usepackage[colorlinks=true,citecolor=blue,linkcolor=blue,urlcolor=blue]{hyperref}
\usepackage[latin9]{inputenc}
\usepackage{hyperref}
\usepackage{color}
\usepackage{float}
\usepackage{textcomp}
\usepackage{amstext}
\usepackage{graphicx}
\usepackage{gensymb}
\usepackage{graphics}\usepackage{subfigure}\usepackage{longtable}\usepackage{pstricks}\usepackage{dcolumn}\usepackage{bm}
\usepackage{siunitx}
\usepackage{ulem}


\usepackage[version=4]{mhchem}

\def\bea{\begin{eqnarray}}
\def\eea{\end{eqnarray}}

\def\vk{{\bf k}}

\begin{document}

\title{Plasmon dispersion in bilayer cuprate superconductors}
\author{M.~Bejas}
\affiliation{Facultad de Ciencias Exactas, Ingenier\'{\i}a y Agrimensura and Instituto de F\'{\i}sica de Rosario (UNR-CONICET), Avenida Pellegrini 250, 2000 Rosario, Argentina}
\author{V.~Zimmermann}
\affiliation{Max Planck Institute for Solid State Research, Heisenbergstra{\ss}e 1, D-70569 Stuttgart, Germany}
\author{D.~Betto}
\affiliation{European Synchrotron Radiation Facility, B.P. 220, 38043 Grenoble, France}
\author{T.~D.~Boyko}
\affiliation{Canadian Light Source, Saskatoon, Saskatchewan S7N 2V3, Canada}
\author{R.~J.~Green}
\affiliation{Department of Physics \& Engineering Physics, University of Saskatchewan, Saskatoon, Saskatchewan, Canada S7N 5A2}
\affiliation{Stewart Blusson Quantum Matter Institute, University of British Columbia, Vancouver, British Columbia V6T 1Z1, Canada}
\author{T.~Loew}
\affiliation{Max Planck Institute for Solid State Research, Heisenbergstra{\ss}e 1, D-70569 Stuttgart, Germany}
\author{N.~B.~Brookes}
\affiliation{European Synchrotron Radiation Facility, B.P. 220, 38043 Grenoble, France}
\author{G.~Cristiani}
\affiliation{Max Planck Institute for Solid State Research, Heisenbergstra{\ss}e 1, D-70569 Stuttgart, Germany}
\author{G.~Logvenov}
\affiliation{Max Planck Institute for Solid State Research, Heisenbergstra{\ss}e 1, D-70569 Stuttgart, Germany}
\author{M.~Minola}
\affiliation{Max Planck Institute for Solid State Research, Heisenbergstra{\ss}e 1, D-70569 Stuttgart, Germany}
\author{B.~Keimer}
\affiliation{Max Planck Institute for Solid State Research, Heisenbergstra{\ss}e 1, D-70569 Stuttgart, Germany}
\author{H.~Yamase}
\email[]{yamase.hiroyuki@nims.go.jp}
\affiliation{Research Center for Materials Nanoarchitectonics (MANA), National Institute for Materials Science (NIMS), Tsukuba 305-0047, Japan}
\author{A.~Greco}
\email[]{agreco@fceia.unr.edu.ar}
\affiliation{Facultad de Ciencias Exactas, Ingenier\'{\i}a y Agrimensura and Instituto de F\'{\i}sica de Rosario (UNR-CONICET), Avenida Pellegrini 250, 2000 Rosario, Argentina}
\author{M.~Hepting}
\email[]{hepting@fkf.mpg.de}
\affiliation{Max Planck Institute for Solid State Research, Heisenbergstra{\ss}e 1, D-70569 Stuttgart, Germany}

\begin{abstract}

The essential building blocks of cuprate superconductors are two-dimensional CuO$_2$ sheets interspersed with charge reservoir layers. In bilayer cuprates, two closely spaced CuO$_2$ sheets are separated by a larger distance from the subsequent pair in the next unit cell.  In contrast to single-layer cuprates, prior theoretical work on bilayer systems has predicted two distinct acoustic plasmon bands for a given out-of-plane momentum transfer. Here we report random phase approximation (RPA) calculations for bilayer systems which corroborate the existence of two distinct plasmons bands. We find that the intensity of the lower-energy band is negligibly small in most parts of the Brillouin zone, whereas the higher-energy band carries significant spectral weight. We also present resonant inelastic x-ray scattering (RIXS) experiments at the O $K$-edge on the bilayer cuprate Y$_{0.85}$Ca$_{0.15}$Ba$_2$Cu$_3$O$_7$ (Ca-YBCO), which show only one dispersive plasmon branch, in agreement with the RPA calculations. In addition, the RPA results indicate that the dispersion of the higher-energy plasmon band in Ca-YBCO is not strictly acoustic, but exhibits a substantial energy gap of approximately 250 meV at the two-dimensional Brillouin zone center.

\end{abstract}

\maketitle

\newpage

\section{Introduction}
Cuprate high-temperature superconductors have garnered a prominent position in modern condensed matter research \cite{Keimer2015}. In addition to superconductivity,  a broad range of phenomena, including the pseudogap, spin and charge density wave orders, and the strange metal phase \cite{Lee2006,Scalapino2012,Armitage2010}, occur when charge carriers are introduced into the CuO$_2$ sheets, which are periodically stacked along the $z$-direction.

The electrodynamics of such a layered configuration draws close parallels to that of the layered electron gas (LEG) model \cite{Grecu1973, fetter1974, Grecu1975}, which was predicted to host unconventional plasmon excitations  \cite{Kresin1988,Bill2003}. Specifically, the plasmon spectrum in a LEG system consists of one optical and several acoustic modes with a characteristic dispersion as a function of both the in-plane ($\textbf{q}_{\parallel}$) and the out-of-plane momentum ($q_z$). For $q_z$ = 0, the plasmon dispersion corresponds to the optical branch for all values of $\textbf{q}_{\parallel}$. For a finite $q_z$, the plasmon dispersion follows the acoustic branch, where the lowest-energy mode is reached when $q_z$ = $\pi$. This distinct three-dimensional (3D) dependence of the plasmon dispersion in a LEG sets it apart from the $\sqrt{q}$ dispersion in a purely two-dimensional (2D) system and the $q^2$ dependence in an isotropic 3D metal.

\begin{figure}[tb]
\begin{centering}
\includegraphics[width=0.9\columnwidth]{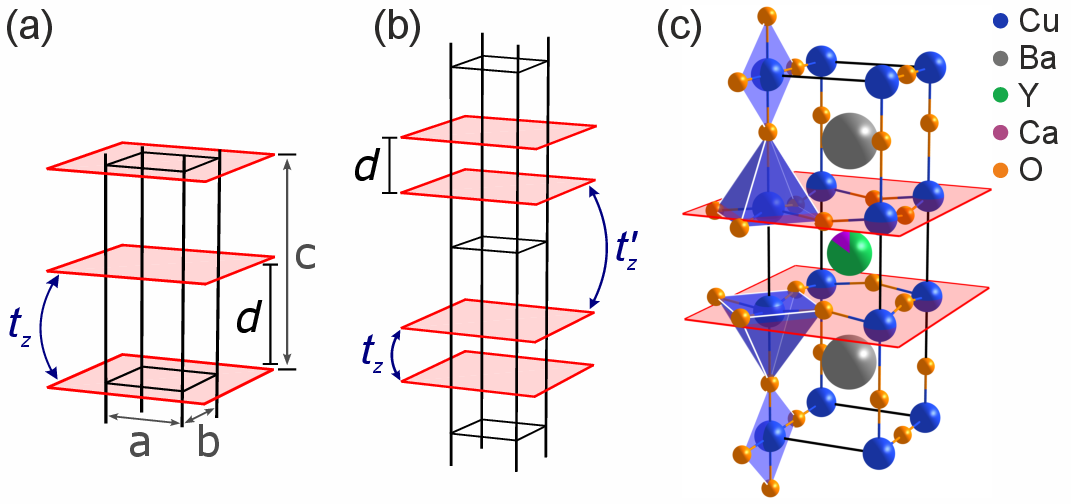}
\par\end{centering}
\caption{(a) Sketch of a system with equally spaced conducting sheets (red shaded planes). Hopping between the sheets is denoted by $t_z$. The black boxes represent unit cells with the lattice parameters $a$, $b$, and $c$. The distance between the planes is denoted by $d$. (b) Sketch of a bilayer system that contains two closely spaced sheets within a unit cell. The intra-bilayer hopping is denoted by $t_z$ and the inter-bilayer hopping by $t_z'$. The distance between the two closely spaced planes is denoted by $d$. (c) Schematic of the crystal structure of the bilayer cuprate Ca-YBCO. The red shaded planes are the CuO$_2$ sheets. 
}
\label{structure}
\end{figure}

To calculate the details of the corresponding plasmon dispersion in cuprates, several computational methods have been employed \cite{Hepting2018,Eremin2018,Fidrysiak2021,Bauer2009,Ruvalds1987,Romero2019,Gabriele2022,Silkin2023}, including the random phase approximation (RPA) \cite{Kresin1988,Grecu1973,Prelovsek1999,Bill2003,Markiewicz2008} and a large-$N$ theory for the layered $t$-$J$ model with long-range Coulomb interaction ($t$-$J$-$V$ model) \cite{Greco2016,Greco2019,Greco2020,Nag2020,Hepting2022}. Yet, the majority of theoretical studies considered cuprates with equidistantly stacked CuO$_2$ sheets  [Fig.~\ref{structure}(a)], whereas the cuprates with the highest superconducting transition temperatures ($T_c$) are multilayer systems \cite{Keimer2015}. Specifically, in a bilayer system, two closely spaced CuO$_2$ planes are separated by a substantially larger distance from the next set of planes in the subsequent unit cell [Fig.~\ref{structure}(b)]. For such bilayer systems, the generalized RPA approach by Griffin and Pindor predicted that for a given $q_z$ the acoustic plasmon modes separate into high- ($\omega_+$) and low-energy ($\omega_-$) bands \cite{Griffin1989}. This is markedly distinct from a system with equidistantly spaced planes [Fig.~\ref{structure}(a)], where one band comprises all acoustic plasmon modes. Furthermore, the bilayer RPA approach \cite{Griffin1989} indicated that, at a given $\textbf{q}_{\parallel}$, the plasmon modes of the $\omega_-$ band are essentially degenerate and do not depend on the value of $q_z$. By contrast, the plasmon modes in the $\omega_+$ band exhibit a $q_z$-dependence that is qualitatively similar to that of systems with equidistant planes.

Similarly to the majority of theoretical studies, preceding resonant inelastic x-ray scattering (RIXS) experiments investigating plasmons in cuprates focused on compounds with equally spaced CuO$_2$ sheets, including ${\rm (La,Nd)_{2-x}Ce_{x}CuO_{4}}$ \cite{Hepting2018,Lin2019,Ishii2019}, ${\rm La_{2-x}Sr_{x}CuO_{4}}$ \cite{Nag2020,Singh2022,Hepting2023,Li2023}, and ${\rm Bi_{2}Sr_{0.16}La_{0.4}CuO_{6+\delta}}$ (Bi-2201) \cite{Nag2020}. RIXS is a particularly versatile tool to study the plasmon phenomenology in cuprates, because it provides high momentum resolution in all three spatial directions along with polarization analysis, which facilitated the identification of acoustic plasmons \cite{Hepting2018,Nag2020,Singh2022}. In addition, recent RIXS experiments revealed that the nominally acoustic plasmon branch in cuprates deviates from a pure acoustic dispersion due to an energy gap at the 2D Brillouin zone (BZ) center \cite{Hepting2022} when charge carrier hopping ($t_z$) between the CuO$_2$ planes is present \cite{Greco2016}. Despite these advances, RIXS studies have yet to explore the plasmon dispersion in multilayer cuprates, including bilayer systems such as ${\rm YBa_{2}Cu_{3}O_{6+\delta}}$ (YBCO) and ${\rm Bi_{2}Sr_{2}CaCu_{2}O_{8+\delta}}$ (Bi-2212). Specifically, the impact of the relatively large interlayer hopping within a bilayer unit on the plasmon spectrum as well as the anticipated coexistence of $\omega_+$ and $\omega_-$ plasmon bands \cite{Griffin1989} have remained open questions, fundamental to a comprehensive description of their electrodynamics. 

In this work, we adapt an RPA framework for bilayer systems beyond that of Ref.~\onlinecite{Griffin1989}, incorporating a cuprate-specific electronic band dispersion and intra-bilayer hopping $t_z$. This enables us to evaluate the relative spectral intensities of plasmon bands in bilayer cuprates. Utilizing RIXS at the O $K$-edge on the bilayer cuprate Y$_{0.85}$Ca$_{0.15}$Ba$_2$Cu$_3$O$_7$ (Ca-YBCO) [Fig.~\ref{structure}(c)], we identify a single dispersive plasmon mode. This observation is compatible with our RPA analysis, which predicts that one of the two plasmon bands carries almost vanishing spectral weight. Furthermore, the experimentally observed plasmon dispersion is captured when using cuprate-typical effective interaction parameters in the RPA, while an extrapolation of the plasmon dispersion to the 2D BZ center suggests the presence of a gap of the acoustic-like branch of approximately 250 meV.

\begin{figure*}[tb]
\begin{centering}
\includegraphics[width=16cm]{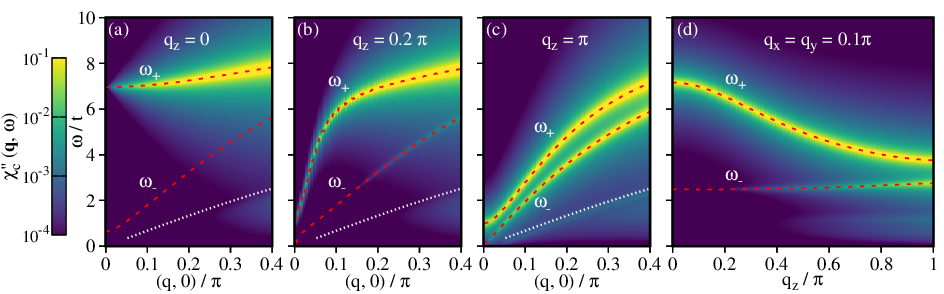}
\par\end{centering}
\caption{Computed intensity maps of the RPA charge response for a generic bilayer system. (a-c) Intensity maps in the ${\bf q}_\parallel$-$\omega$ plane for $q_z=0$, $0.2\pi$, and $\pi$, respectively. Red dashed lines correspond to the  zeros of the dielectric function ($\omega_+$ and $\omega_-$). White dotted lines are guides to the eye marking the upper bound above which the spectral weight of the particle-hole continuum becomes insignificant. (d) Intensity map in the $q_z$-$\omega$ plane for $q_x=q_y=0.1\pi$.}
\label{Griffin}
\end{figure*}

\section{Results}
For the RPA calculations, we consider a bilayer setting [Fig.~\ref{structure}(b)] under the presence of a long range Coulomb interaction. In order to extract fundamental characteristics of bilayer systems without compromising generality, we start with a generic model. The Hamiltonian is given by $H=H_0+H_I$, with the single particle Hamiltonian 

\begin{equation} \label{rashba_hubbard}
H_0 = \sum_{{\bf k},\sigma,\alpha,\beta} {c}^\dag_{{\bf k},\sigma,\alpha}  \hat{h}_{\alpha \beta} ( {\bf k} ) c^{\ }_{{\bf k},\sigma,\beta}  
\end{equation}

\noindent where $\hat{h}_{\alpha \beta}$ are the elements of the matrix
\begin{equation}
\hat{h}({\bf k} ) = 
\left(
\begin{array}{ll}
\varepsilon^\parallel_{\vk} & \varepsilon^\perp_{\vk}\\
\varepsilon^{\perp \, *}_{\vk} & \varepsilon^\parallel_{\vk}
\end{array}
\right).
\label{singleH}
\end{equation}

\noindent Here $\varepsilon^\parallel$ is the in-plane dispersion, 

\begin{equation}\label{eq_disp_in_plane}
\varepsilon^\parallel_{\vk}= - 2t(\cos k_x+\cos k_y),
\end{equation}

\noindent which contains the nearest neighbor hopping $t$, and $\varepsilon^\perp$ is the out-of-plane dispersion,

\begin{equation}\label{eq_disp_out_plane}
\varepsilon^\perp_{\vk}= - t_z e^{ik_zd},
\end{equation}

\noindent which contains only the intra-bilayer hopping $t_{z}$, while omitting the inter-bilayer hopping $t'_{z}$ [Fig.~\ref{structure}(b)]. The electron annihilation operator is denoted by $c_{{\bf k},\sigma,\alpha}$, where $\sigma$ is the spin index and $\alpha, \beta=1,2$ indicates each plane in the unit cell.
The symbol '$*$' means complex conjugate.  In the following, the in-plane hopping $t$ is considered as the energy unit.   

The Coulomb interacting Hamiltonian is given by 

\begin{equation} \label{rashba_hubbard}
H_I = \sum_{{\bf q},\alpha,\beta} n_\alpha({\bf q})  V_{\alpha \beta}({\bf q}) n_\beta({\bf -q}), 
\end{equation}

\noindent where $n_\alpha({\bf q})$ is the electron density at a given plane $\alpha$, and the two forms of the Coulomb interactions are

\begin{equation}
V_{\alpha \alpha}({\bf q})=\frac{V_c}{|{\bf q_\parallel}|}
\left(\frac{\sinh|{\bf q_\parallel}|c}{\cosh|{\bf q_\parallel}|c-\cos q_zc}\right),
\end{equation}

\noindent and

\begin{equation}
V_{\alpha \beta}({\bf q})=\frac{V_c}{|{\bf q_\parallel}|}
\left(\frac{\sinh|{\bf q_\parallel}|(c-d)+e^{-\mathrm{i}q_zc}    \sinh|{\bf q_\parallel}|d}{\cosh|{\bf q_\parallel}|c-\cos q_zc}\right)e^{-\mathrm{i}q_zd},
\end{equation}

\noindent with $V_{\alpha \alpha}=V_{\beta \beta}$ and $V_{\alpha \beta}=V^*_{\beta \alpha}$ ($\alpha \neq \beta$)~\cite{Griffin1989}. Note that, in contrast to a simplified treatment of the conducting sheets in a 2D LEG picture in Ref.~\cite{Griffin1989}, our approach  incorporates tight-binding dispersions [see Eqs.~\ref{eq_disp_in_plane} and \ref{eq_disp_out_plane}], which allow for the accommodation of bandstructure effects.

Within RPA, the dressed charge susceptibility $\hat{\chi}({\bf q},i\omega_l)$ 
is given by
\begin{equation} \label{eq_RPA_suscep}
\hat{\chi} ({\bf q},i\omega_l)
=
\left[ I -  \hat{\chi}^{(0)} ( {\bf q}, i \omega_l ) \hat{V}({\bf q})    \right]^{-1} 
\hat{\chi}^{(0)} ({\bf q},i\omega_l), 
\end{equation}
where $\hat{\chi}^{(0)}$ is the bare charge  susceptibility. Here, $\hat{\chi}^{(0)}$ and $\hat{\chi}$ are $2 \times 2$ matrices. The dielectric function $\epsilon({\bf q},i\omega_l)$ can be calculated as the determinant of the matrix $[ I - \hat{\chi}^{(0)}  \hat{V} ]$. In the above expressions, the bare susceptibility $\hat{\chi}^{(0)}$
is defined as the convolution of two Green's functions

\begin{subequations} \label{def_chi_zero}
\begin{align}
\chi^{(0)}_{\alpha \beta} (\mathbf{q},i\omega_l) &= \frac{2T}{N_s}\sum_{\mathbf{k}, i\nu_n} G^{(0)}_{\alpha \beta}  (\mathbf{k},i\nu_n) G^{(0)}_{\beta \alpha}  (\mathbf{k}+ \mathbf{q},i\nu_n+i\omega_l), 
\label{chi0}
\end{align}
with  
\begin{equation}
\hat{G}^{(0)}( \mathbf{k} , i \nu_n ) = 
\left(
\begin{array}{ll}
i \nu_n-(\varepsilon^\parallel_{\vk}-\mu)&-\varepsilon^\perp_{\vk}\\
-\varepsilon^{\perp \, *}_{\vk} & i \nu_n-(\varepsilon^\parallel_{\vk}-\mu)
\end{array}
\right)^{-1}
\label{2x2Green}
\end{equation}
\end{subequations}

\noindent the  $2 \times 2$ bare electronic Green's function.
$\omega_l$ ($\nu_n$) is a  bosonic (fermionic)  
Matsubara frequency. For the momentum ${\bf q_\parallel}$ ($q_z$), we use the units of the in-plane (out-of-plane) lattice constant $a$ ($c$). $T$ ($N_s$) is the temperature (number of sites). 

We compute the $2\times2$ dressed RPA charge susceptibility, and execute the analytical continuation $\mathrm{i}\omega_n=\omega+\mathrm{i} \Gamma$, where $\Gamma$ is in principle infinitesimally small. We then take the imaginary part ${\chi''}_{\alpha \beta}({\bf q},\omega)=-{\rm Im} {\chi}_{\alpha \beta}({\bf q},\omega+\mathrm{i}\Gamma)$ of the charge susceptibility. The full charge response $\chi''_c({\bf q},\omega)$ is

\begin{equation}
\chi''_c({\bf q},\omega)=\sum_{\alpha,\beta} \chi''_{\alpha \beta}({\bf q},\omega).
\end{equation}

We employ the following generic parameter set: $c=3a$ and $d=c/3$ for the structural parameters, $t_z/t=0.1$ for the hopping, and the temperature $T=0$. The chemical potential $\mu$ and the broadening $\Gamma$ are set to $\mu/t=-1$ and $\Gamma/t=0.1$, respectively. 
The Coulomb repulsion strength is parameterized by $V_c$. For the purpose of illustration, we use the value $V_c/t=50$, which allows us to emphasize the intensity of both $\omega_+$ and $\omega_-$ modes.

Figure~\ref{Griffin} shows the computed intensity maps of the charge response, i.e., the imaginary part of the charge susceptibility. Panels (a)-(c) present results within the ${\bf q_\parallel}$-$\omega$ plane for $q_z=0$, $0.2\pi$, and $\pi$, respectively. Red dashed lines in each panel indicate the two zeros of the dielectric function, $\omega_+$ and $\omega_-$, which are situated above the particle-hole continuum (white dotted lines).  Note that a logarithmic scale was used for the intensities in the maps to accentuate the contrast between the spectral weights of $\omega_+$, $\omega_-$, and the particle-hole continuum. Interestingly, although both modes are zeros of the dielectric function, the spectral weight of the $\omega_-$ mode is much lower than that of the $\omega_+$ mode. In particular, for $q_z=0$, the spectral weight of $\omega_-$  is exactly zero [Fig.~\ref{Griffin}(a)] for all $\bf q_\parallel$ values. Conversely, $\omega_+$ corresponds to the conventional optical plasmon branch, experiencing a decline in spectral intensity as ${\bf q_\parallel}$ decreases, reaching exactly zero intensity at ${\bf q_\parallel}=(0,0)$. For $q_z=0.2\pi$, some subtle spectral weight emerges for $\omega_-$ at large ${\bf q_\parallel}$ values [Fig.~\ref{Griffin}(b)]. For $q_z=\pi$, the spectral weight of $\omega_-$ becomes more discernible across the entire ${\bf q_\parallel}$ range [Fig.~\ref{Griffin}(c)]. In addition, we note that for finite $q_z$ values [Figs.~\ref{Griffin}(b),(c)], an energy gap is present for the $\omega_+$ band at ${\bf q_\parallel}=(0,0)$, whereas $\omega_-$ approaches zero-energy. Hence, in contrast to the purely acoustic plasmon bands in Ref.\cite{Griffin1989}, we find that a gap opens for the $\omega_+$ band as a direct consequence of the inclusion of a finite $t_z$ in our model. 

For a more detailed analysis of the evolution of the plasmon spectral weight along the out-of-plane direction, we plot the intensity map within the $q_z$-$\omega$ plane for $q_x=q_y=0.1\pi$ in Fig.~\ref{Griffin}(d). Notably, the dispersion of $\omega_-$ appears to be almost independent of $q_z$, whereas the behavior of $\omega_+$ with a maximum at $q_z = 0$ and a minimum at $q_z = \pi$ is reminiscent of that of plasmons in systems with equidistant sheets \cite{Hepting2018,Greco19,Nag2020}. In addition, we find that the spectral weight of $\omega_-$ decreases strongly with decreasing $q_z$ and is nearly invisible for small $q_z$. In comparison, in layered systems with equidistantly stacked planes, such as single-layer cuprates, the $q_z = \pi$ acoustic branch, corresponding to an out-of-phase charge distribution, does not exhibit zero intensity \cite{Hepting2018,Nag2020,Greco19}. Conversely, a configuration comprising solely two isolated planes is characterized by the presence of only two plasmon branches,  corresponding to in-phase and out-of-phase charge distributions \cite{santoro88}, with the latter exhibiting zero spectral intensity. Systems with periodically stacked bilayers, such as bilayer cuprates, present a combination of elements from both previous cases. As a result, the $\omega_-$ mode and generally configurations with partial out-of-phase charge distributions manifest diminished but non-zero spectral intensities, and zero-intensity at $q_z = 0$.

To assess the predictive power of our theoretical model, we perform RIXS experiments  in resonance to the O $K$-edge, known to provide significant plasmon intensity for hole-doped cuprates \cite{Nag2020,Hepting2023,Singh2022}. Specifically, we carry out RIXS measurements on a detwinned Ca-YBCO single-crystal ($p = 0.21$) \cite{Minola2015} with a superconducting transition temperature $T_c \approx 75$ K. In addition to this overdoped variant of YBCO, we also conduct O $K$-edge RIXS measurements on an optimally doped YBCO film (without Ca substitution), which yield qualitatively comparable results (see Appendix~\ref{app:YBCO}). 

Figure~\ref{structure}c shows the crystallographic unit cell of Ca-YBCO, with the lattice constants $a = 3.89$ {\AA}, $b = 3.88$ {\AA}, and $c = 11.68$ {\AA}. The spacing between the two CuO$_2$ planes (intra-bilayer spacing) is $d = 3.36$ {\AA}, and the distance from the center of the bilayer to the center of the next pair of planes along the $c$-axis direction (inter-bilayer spacing) is 11.68 {\AA} (i.e., equivalent to the $c$-axis lattice constant).
The RIXS spectra were collected with high energy resolution ($\Delta E \approx$ 27 meV) at $T = 20$ K at the ID32 beamline of the ESRF \cite{Brookes2018}. A similar scattering geometry as in Ref.~\onlinecite{Nag2020} was employed, with the $b$-axis and the $c$-axis of Ca-YBCO lying in the scattering plane and incident photons linearly polarized perpendicular to the scattering plane ($\sigma$-polarization). The scattering angle was varied by continuous rotation of the RIXS spectrometer arm, which in combination with a rotation of the sample enabled for the variation of the in-plane ($\textbf{q}_{\parallel}$) and out-of-plane momentum transfer ($q_{z}$) independently of each other.

\begin{figure}[tb]
\begin{centering}
\includegraphics[width=0.8\columnwidth]{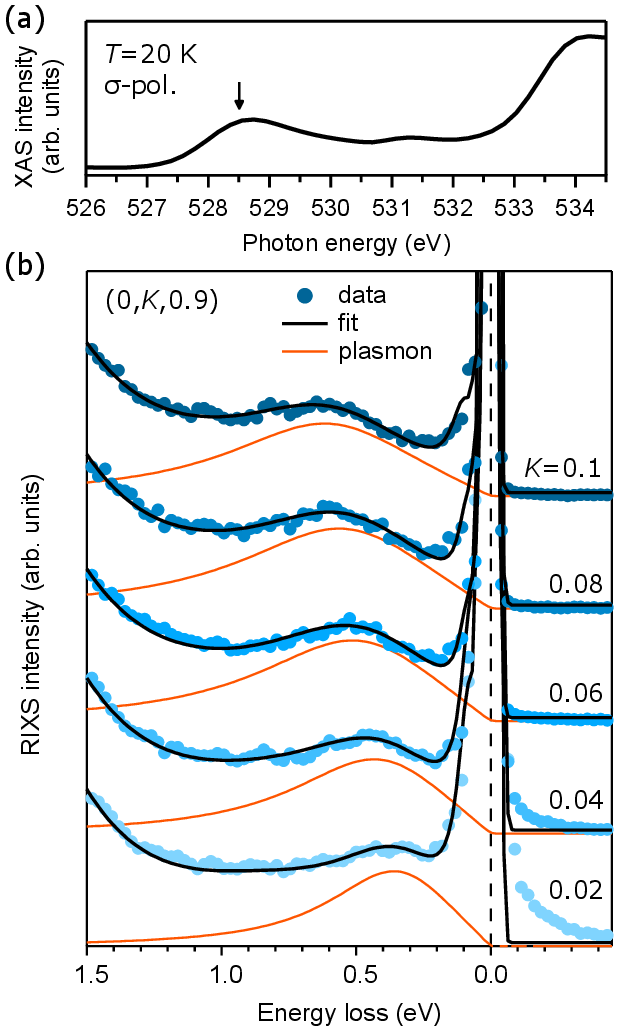}
\par\end{centering}
\caption{(a) XAS of Ca-YBCO across the O $K$-edge measured with $\sigma$-polarized photons. The arrow indicates the incident photon energy (528.5 eV) used for the RIXS experiment. (b) RIXS spectra of Ca-YBCO measured at various momenta along the $K$ direction, while $H$ and $L$ were fixed to 0 and 0.9, respectively. The fit (black line) to the experimental
data (blue symbols) includes the plasmon peak (orange peak
profile) and other contributions (not shown here) that are described in Appendix~\ref{app:RIXS}. Curves for different momenta are offset in the vertical direction for clarity. 
}
\label{RIXS}
\end{figure}

Figure~\ref{RIXS}a shows the XAS signal of Ca-YBCO across the near-edge fine structure of the O $K$-edge. A prominent peak feature emerges around 528.5 eV, similarly to the hole-peak reported in the O $K$-edge XAS of various other doped cuprates \cite{Chen1991,Chen1992,Pellegrin1993,Brookes2015,Ishii2017}. The hole-peak in cuprates is associated with the Zhang-Rice singlet states, corresponding to plaquettes of hybridized states between copper and oxygen ions. In previous RIXS experiments on hole-doped cuprates with equidistant CuO$_2$ sheets, dispersive plasmon excitations were detected for incident photon energies coinciding with the energy of the hole-peak in the O $K$-edge XAS \cite{Nag2020,Singh2022,Hepting2023}.

Figure~\ref{RIXS}b presents the RIXS spectra for various momentum transfers, acquired with an incident photon energy of 528.5 eV. The momentum transfer is denoted by $(H,K,L)$ in reciprocal lattice units $(2\pi/a, 2\pi/b, 2\pi/c)$. Throughout the experiment, the out-of-plane momentum is fixed to $L = 0.9$, while the in-plane momentum transfer is varied along the $K$ direction, although we note that a qualitatively similar plasmon dispersion can be expected for momentum transfer along the $H$ direction of Ca-YBCO. We fit the RIXS spectra by the sum of the elastic line at zero-energy loss and several damped harmonic oscillator functions for the inelastic features. In addition, intense peaks from fluorescence and $dd$-excitations emerge beyond 2 eV energy loss, whose tails extend down to 1 eV and below, and are captured by an exponential function in our fits. Further details of the fitting procedure are given in Appendix~\ref{app:RIXS}.

Notably, the RIXS spectra exhibit a peak that disperses approximately from 350 to 600 meV when the momentum component $K$ increases from 0.02 to 0.1 [Fig.~\ref{RIXS}b]. Such a rapid dispersion within a small variation of the in-plane momentum is reminiscent of that of plasmons in other cuprates \cite{Hepting2018,Hepting2022,Lin2019,Nag2020,Singh2022,Ishii2019}. In contrast, magnetic excitiations in doped cuprates, such as bi-paramagnons, were found to be (almost) non-dispersive in O $K$-edge RIXS \cite{Bisogni2012a,Bisogni2012b,Nag2020,Singh2022}, and the bandwidth of the dispersion of paramagnons in Cu $L$-edge RIXS is typically less than 300 meV. Moreover, the energy scale of the dispersive peak in Fig.~\ref{RIXS}b is situated far above that of phonons in cuprates, which is typically below 100 meV. Hence, we tentatively  assign the observed feature to a plasmon. 

At first sight, our observation of a single dispersive plasmon mode in bilayer Ca-YBCO appears to conflict with the emergence of $\omega_+$ and $\omega_-$ bands on equal footing \cite{Griffin1989}, but might be compatible with our RPA model predicting  a vanishingly small spectral weight of the $\omega_-$ band. Thus, for further scrutiny, we next apply our bilayer RPA approach to fit the plasmon dispersion observed in the RIXS experiment on Ca-YBCO. To this end, we  include the next-nearest neighbor hopping $t'$ in the in-plane dispersion  $\varepsilon^\parallel_{\vk}= - 2t(\cos k_x+\cos k_y)-4t'\cos k_x \cos k_y$, such that $t'/t=-0.3$ with $t=0.35$ eV, as discussed in Ref.~[\onlinecite{Andersen1995}] for YBCO. Furthermore, we choose $V_{c}/t =25$ and $t_z/t=0.06$, along with the out-of-plane dispersion $\varepsilon^\perp_{\vk}= -t_z (\cos(k_x)-\cos(k_y))^2e^{ik_zd}$, which was proposed for YBCO in Ref.~[\onlinecite{Andersen1995}]. In cuprates,  inter-bilayer hopping [Fig.~\ref{structure}(b)] is typically very small ($t'_{z} \ll t_{z}$) \cite{Andersen1995}, and is therefore not considered in the following. The charge carrier doping is set to $\delta=0.21$, in accord with the Ca-YBCO sample of the experiment. Since YBCO is a strongly correlated system, correlations are expected to affect the bare parameters. As discussed in Ref.~[\onlinecite{Greco2016}], taking into account these renormalization effects is important for an accurate description of the plasmon dispersions reported in RIXS experiments on cuprates \cite{Nag2020,Hepting2022,Hepting2023,Greco2020}, prompting only the renormalization of the bare parameters $t, t'$, and $t_z$ by a factor $\delta$. Note that the slope and the bandwidth of the computed plasmon dispersion are generally not dictated by a single parameter alone, but result from a combination of the hopping parameters and the parameter $V_c$.

\begin{figure}[tb]
\begin{centering}
\includegraphics[width=1\columnwidth]{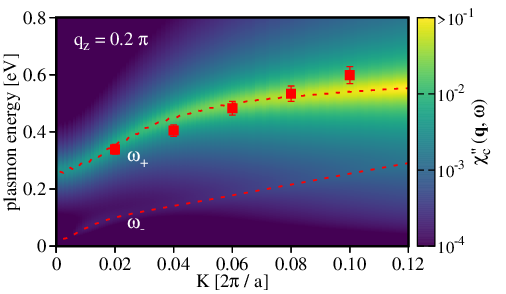}
\par\end{centering}
\caption{Computed intensity maps of the RPA charge response using renormalized effective parameters optimized for the bilayer cuprate Ca-YBCO. The dispersion is along the $K$-direction for $q_z = 0.2\pi$. The superimposed red symbols are the plasmon energies for Ca-YBCO extracted from fits to the RIXS data [see Fig.~\ref{RIXS}(b)].}
\label{theory}
\end{figure}

Figure~\ref{theory} shows the computed intensity map along the $K$-direction for $q_z = 0.2\pi$ together with the experimentally extracted plasmon energies (red symbols) at $L=0.9$ ($q_z=1.8\pi$). We assume that this out-of-plane momentum employed in the experiment mirrors the essential features of the plasmon dispersion at $q_z = 0.2\pi$ in the first BZ. For completeness, we also present an intensity map computed for bare Ca-YBCO parameters without renormalization in Appendix~\ref{app:RPA}, yielding qualitatively similar results. The dispersion of the $\omega_+$ mode (upper red dashed line) in Fig.~\ref{theory} agrees remarkably well with the RIXS experiment, corroborating our previous assignment of the dispersive mode to a plasmon excitation. At the 2D BZ center, the computed $\omega_+$ dispersion exhibits an energy gap of approximately 250 meV. In contrast, the computed dispersion of the $\omega_-$ mode (bottom red dashed line) is essentially gapless, and its marginal spectral weight is almost indiscernible in the color map in spite of a logarithmic intensity scale. This suggests that at least for $L=0.9$ the $\omega_-$ mode is below the detection limit of RIXS experiments, whereas the $\omega_+$ mode can be prominently observed.

\section{Discussion and conclusion}
The results of our study have several implications. First, they underscore the distinct nature of charge excitations in bilayer systems, as already pointed out in earlier works investigating the zeros of the dielectric function \cite{Griffin1989,santoro88}. However, while our RPA calculations corroborate that the emergence of $\omega_+$ and $\omega_-$ bands is a general characteristic of bilayer systems, we have revealed that the spectral weight of the $\omega_-$ band is negligibly small throughout most parts of the BZ. This also rationalizes why only one dispersive plasmon band ($\omega_+$) is observed in our RIXS data on bilayer Ca-YBCO.

Another insight from our study is the remarkably large plasmon gap of about $250$ meV at the center of the BZ.  
This value of the gap in bilayer Ca-YBCO is possibly attributable to the large intra-bilayer hopping $t_z/t=0.06$, which is consistent with the magnitude of the hopping discussed in Ref.~\cite{chakravarty93}. 

Despite our model's ability to quantitatively and qualitatively describe the observed plasmon mode in the bilayer cuprate Ca-YBCO, there remain several open areas for future research. For instance, although the predicted spectral weight of the $\omega_-$ band is diminishing, it might be detectable with RIXS for momenta close to $q_z = \pi$ and for large ${\bf q_\parallel}$ values, where its maximum intensity is expected  according to Figs.~\ref{Griffin}(c,d). A telltale signature to discern between the $\omega_+$ and $\omega_-$ bands in this part of the BZ might be their dispersive versus non-dispersive character in a RIXS scan along the $q_z$ direction. Furthermore, since cuprates are strongly correlated systems, implementing the comprehensive $t$-$J$-$V$ model calculation for the bilayer lattice including the intra- and inter-layer hoppings should be a consideration for future work.

\begin{acknowledgements}
We thank C.~Falter for fruitful discussions and A.~P. Schnyder for critical reading of the manuscript. A.~G. acknowledges the Max Planck Institute for Solid State Research in Stuttgart for hospitality and financial support. H.~Y. was supported by JSPS KAKENHI Grant No. JP20H01856 and World Premier International  Research Center Initiative (WPI), MEXT, Japan. Parts of the results presented in this work were obtained by using the facilities of the CCT-Rosario Computational Center, member of the High Performance Computing National System (SNCAD, MincyT-Argentina). Part of the research described in this
paper was performed at the Canadian Light Source, a national research facility of the University of Saskatchewan, which is supported by the Canada Foundation for Innovation (CFI), the Natural Sciences and Engineering Research Council (NSERC), the National Research Council (NRC), the Canadian Institutes of Health Research (CIHR), the Government of Saskatchewan, and the University of Saskatchewan.
\end{acknowledgements}

\bibliography{cuprates}

\newpage

\appendix

\section{Complementary RIXS measurements}
\label{app:YBCO}

\begin{figure}[tb]
\begin{centering}
\includegraphics[width=0.8\columnwidth]{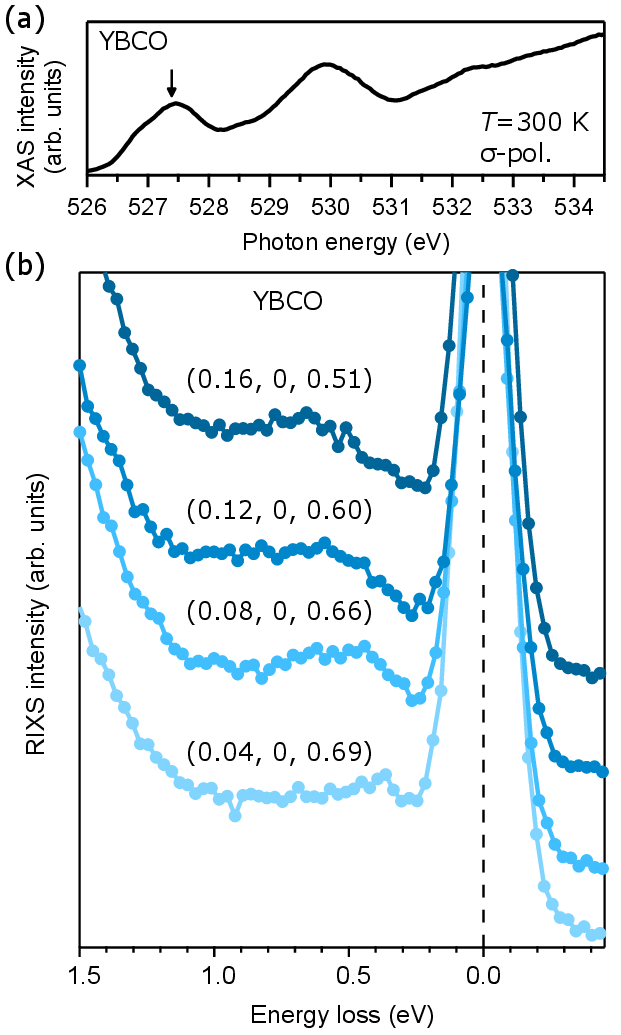}
\par\end{centering}
\caption{(a) XAS of the YBCO film across the O $K$-edge. The arrow indicates the incident photon energy (527.32 eV) used for the RIXS experiment. (b) RIXS spectra of the YBCO film measured at various momenta along the $H$ direction between $H = 0.04$ and 0.16. Curves for different momenta are offset in the vertical direction for clarity. 
}
\label{app:FigSUP}
\end{figure}

In addition to the RIXS experiment on the Ca-YBCO single crystal, we investigated plasmon excitations in an YBCO thin film. The epitaxial film with a thickness of approximately 50 nm was grown by pulsed laser deposition (PLD) on a (001) oriented SrTiO$_3$ substrate. The lattice parameters were $a,b = 3.87$ {\AA} and  $c = 11.72$ {\AA}. After the growth, the YBCO film was annealed in oxygen atmosphere in order to achieve full oxygenation, corresponding to a hole-doping of $p = 0.19$. The measured $T_c$ was 83 K. 

The O $K$ edge RIXS measurements were performed at the REIXS beamline of the Canadian Light Source (CLS). The XAS data in Fig.~\ref{app:FigSUP}(a) were taken with $\sigma$-polarized photons at an incident angle $\theta = 35^\circ$ in partial fluorescence yield using a silicon drift detector, collecting only the O $K_\alpha$ emission line. The RIXS spectra in Fig.~\ref{app:FigSUP}(b) were collected at 300 K using a Rowland circle spectrometer with a combined energy resolution $\Delta E \sim 190$ meV and  linearly polarized photons ($\sigma$-polarization). The $c$ and the $a/b$-axes of the twinned film were lying in the scattering plane. The scattering angle was kept fixed at $90^\circ$, while the angle $\theta$ between the sample surface and the incident x-rays was varied for momentum dependent measurements. In this scattering configuration, a variation of the in-plane momentum transfer $\textbf{q}_{\parallel}$ also leads to a (small) change of the out-of-plane momentum $q_z$. This variation is different from the experiment on Ca-YBCO in the main text, where a variation of the in-plane ($\textbf{q}_{\parallel}$) and out-of-plane momentum transfer ($q_{z}$) independently of each other allowed us to fix the out-of-plane momentum to $L=0.9$.

Figure~\ref{app:FigSUP}(b) shows the RIXS spectra of the YBCO film measured at various momenta along the $H$ direction, normalized to the incident flux. The spectra were acquired with an incident photon energy of 527.32 eV, corresponding to the energy of the pre-peak in the XAS data [Fig.~\ref{app:FigSUP}(a)]. The change of $H$ from 0.04 to 0.16 involves a concomitant change of $L$ from 0.69 to 0.51. Hence, the measured plasmon dispersion in the YBCO film is not directly comparable to that of the Ca-YBCO crystal in Fig.~3(b) of the main text, where the out-of-plane momentum was fixed to $L = 0.9$. Nevertheless, the dispersive character and the energy scale of the dispersion of the plasmon in both samples are similar. Clear signatures of the $\omega_-$ modes are not observed in Fig.~\ref{app:FigSUP}(b), even for momenta close to $q_z = \pi$. This absence is consistent with expectations according to  our calculation for Ca-YBCO [Fig.~\ref{theory}], where $V_c/t=25$ was utilized, resulting in a diminishing intensity for the $\omega_-$ mode, even for momenta near $q_z = \pi$. However, it is possible that a weak $\omega_-$ mode signal may be overshadowed by the strong background signal in the O $K$-edge RIXS spectra shown in Fig.~\ref{app:FigSUP}(b). Consequently, this underscores the necessity for future high-resolution RIXS experiments, particularly focused on the vicinity of $q_z = \pi$, to uncover the subtle indications of the $\omega_-$ mode.

\section{RIXS raw data and fits}
\label{app:RIXS}

\begin{figure}
 \begin{centering}
\includegraphics[width=.9\columnwidth]{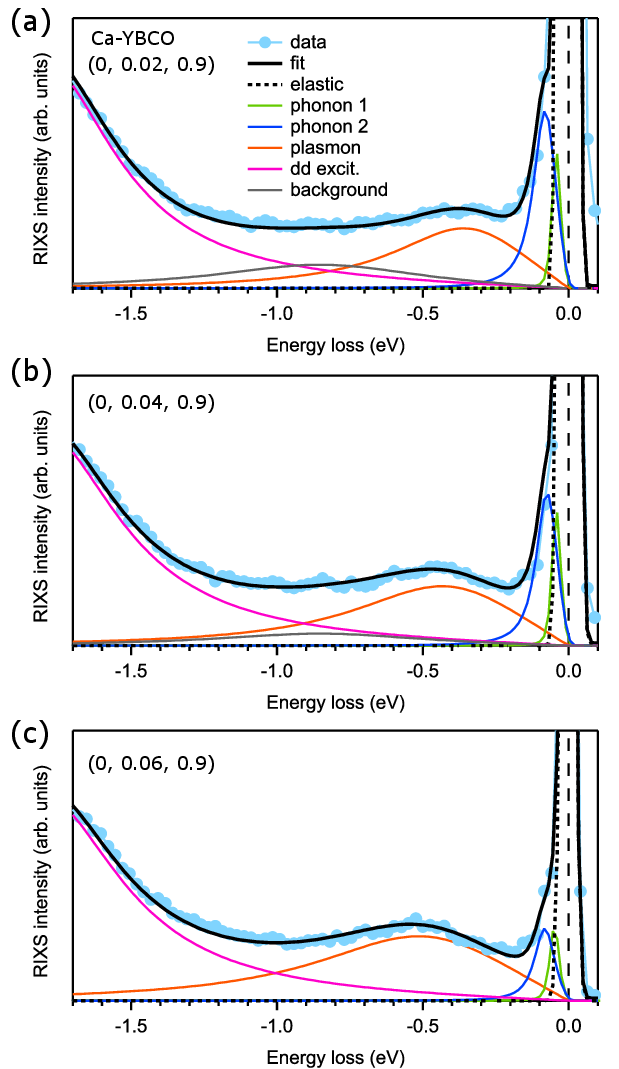}
\par\end{centering}
\caption{Representative fits of the RIXS data. (a) Fit of the RIXS spectrum for momentum (0, 0.02, 0.9), with a model using a Gaussian for the elastic peak (dashed black line) and anti-symmetrized Lorentzians for the other contributions, convoluted with the energy resolution of 27 meV via Gaussian convolution. The individual contributions are described in the text. (b,c) Fit of the spectrum for momentum (0, 0.04, 0.9) and (0, 0.06, 0.9), respectively.}
\label{fig:fits}
\end{figure}

Figure~\ref{fig:fits} displays representative fits of RIXS spectra of the Ca-YBCO single crystal measured at the ID32 beamline of the ESRF. The components fitted to the spectrum are the elastic peak modeled by a Gaussian, and the other contributions are modeled by anti-symmetrized Lorentzians \cite{Hepting2018,Nag2020}, convoluted with the energy resolution of 27 meV via Gaussian convolution. The anti-symmetrized Lorentzian profiles ensure zero intensity at zero energy loss (prior to convolution) for the inelastic features. In detail, the inelastic features are assigned to two phonons, a plasmon, and $dd$-excitations. We note that in our fits the linewidth of the lowest-energy inelastic feature (denoted as phonon 1) is likely underestimated, due to overlap with the highly intense elastic line. For the $dd$-excitations, only the low-energy tail below 1.7 eV energy loss was fitted. For spectra with small in-plane momentum transfer, an additional phenomenological background contribution was fitted. This broad and almost featureless background is reminiscent of the background observed in the low-energy region of RIXS spectra of other cuprates \cite{Hepting2018,Fumagalli2019,Hepting2022,Hepting2023} and possibly originates from charge excitations with (partly) incoherent character and/or a bimagnon continuum \cite{Minola2015}.

\section{RPA calculation using bare parameters}
\label{app:RPA}

\begin{figure}[b]
\begin{centering}
\includegraphics[width=0.9\columnwidth]{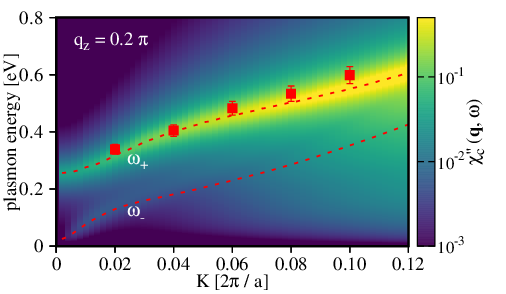}
\par\end{centering}
\caption{Computed intensity map in analogy to Fig.~\ref{theory}, but for bare parameters of Ca-YBCO}
\label{theory_bare}
\end{figure}

In the main text, the RPA results for the bilayer system using the renormalized Ca-YBCO parameters are shown. For comparison, we present the results for the bare parameters of Ca-YBCO in this Appendix [Fig.~\ref{theory_bare}], using $V_c/t=2.5$.

\end{document}